\documentclass[RNAAS]{aastex631}
\usepackage{natbib}
\usepackage{amsmath}
\usepackage{graphicx}
\usepackage{subfig}
\begin{document}
\title{LOFAR search for radio emission from galaxies hosting tidal disruption events}

\author{Kamakshi Kaushik}
\affiliation{Interdisciplinary School of Sciences \\
Savitribai Phule Pune University \\
Pune, Maharashtra, 411007, India}

\author{Amitesh Omar}
\affiliation{Aryabhatta Research Institute of Observational Sciences,\\
Manora Peak, Nainital,\\
Uttrakhand, 263001, India}

\author{Brijesh Kumar}
\affiliation{Aryabhatta Research Institute of Observational Sciences,\\
Manora Peak, Nainital,\\
Uttrakhand, 263001, India}

\author{Kuntal Mishra}
\affiliation{Aryabhatta Research Institute of Observational Sciences,\\
Manora Peak, Nainital,\\
Uttrakhand, 263001, India}

\author{Jagdish Joshi}
\affiliation{Aryabhatta Research Institute of Observational Sciences,\\
Manora Peak, Nainital,\\
Uttrakhand, 263001, India}

\begin{abstract}
Radio emission from 23 tidal disruption event host galaxies were searched in the 144 MHz LOFAR-LoTSS2 images. Three host galaxies are detected with diffuse radio emission, which can be interpreted as either galactic synchrotron radio emission or diffuse radio halo in dense galactic environments. Non-detection of (transient) radio emission in majority of galaxies could be due to self-absorption of radio emission associated with the tidal disruption event.   
\end{abstract}

\section{Introduction} \label{sec:intro}

A total of 56 tidal disruption events (TDEs) are catalogued in \cite{2021ARA&A..59...21G}. A star of mass $M_{*}$ and radius $R_{*}$ upon passing within the tidal radius $R_{*}(M_{BH}/M_{*})^{1/3}$ of a black hole (BH) with mass $M_{BH}$, is tidally ripped apart and the event is known as a TDE \citep{1988Natur.333..523R}. A TDE can result in to a slowly varying transient electromagnetic signal from X-rays to radio bands on time-scales of years. The radio follow-ups of TDEs are discussed in \cite{2020SSRv..216...81A}. Due to self-absorption, radio spectrum shows a turnover around a few GHz, where the emission peaks.  The emission decreases at both higher and lower frequencies from the peak frequency. The radio spectrum evolves over several years and turnover frequency is expected to shift at lower frequencies although with a decrement in the total radio flux \citep{2021ApJ...908..125C}. This makes detection of radio emission from TDEs at low radio frequencies extremely difficult with the present generation of radio telescopes, and may only be possible at very late stages of evolution in some very bright TDEs. 

Studying the properties of host galaxies is important in understanding TDEs. It was found that TDEs are hosted dominantly in post starburst galaxies, where the current star formation rate is very low \citep{2017ApJ...835..176F}. This makes host galaxies weak radio sources. In a study of a sample of 513 PSB galaxies, only 12 galaxies were detected in radio emission above 1 mJy using the 1.4 GHz VLA FIRST images \citep{2011MNRAS.410.1583S}. At lower radio frequencies, diffuse galactic synchrotron radio emission is expected to be enhanced.  Radio detection from TDE host galaxies at low radio frequency (e.g., 144 MHz) can be important for several reasons, e.g., persistent radio continuum from host galaxy can be detected and subtracted to get radio light curves of TDEs, radio morphology and properties of the TDE host galaxies can be studied etc. 

\section{LOFAR detections}

\begin{figure}
\subfloat[{\bf AT2018iih} (0.1 mJy beam$^{-1}\times(1,2,3,4)$)\label{fig:tde1}]
  {\includegraphics[width=.33\linewidth]{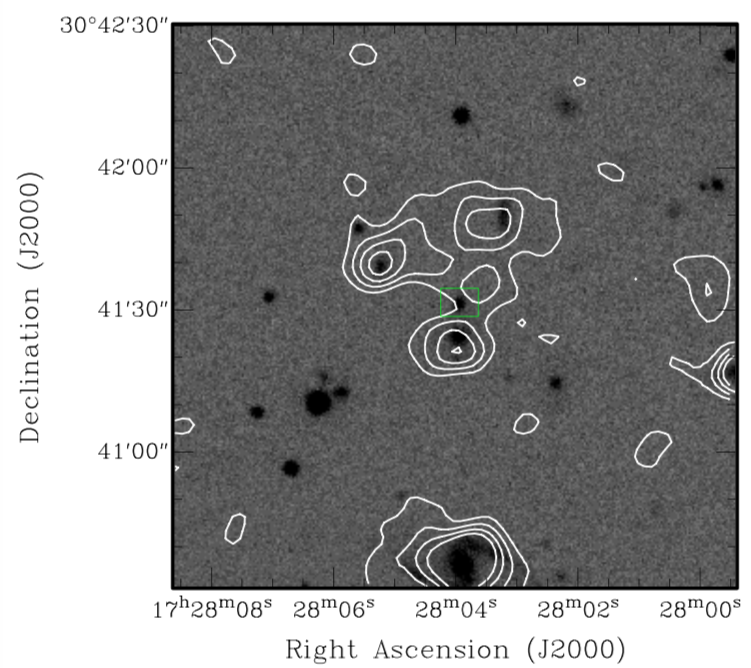}}\hfill
\subfloat[{\bf RBS 1032} (0.04 mJy beam$^{-1}\times(1,2,3,...,9)$)\label{fig:tde2}]
  {\includegraphics[width=.33\linewidth]{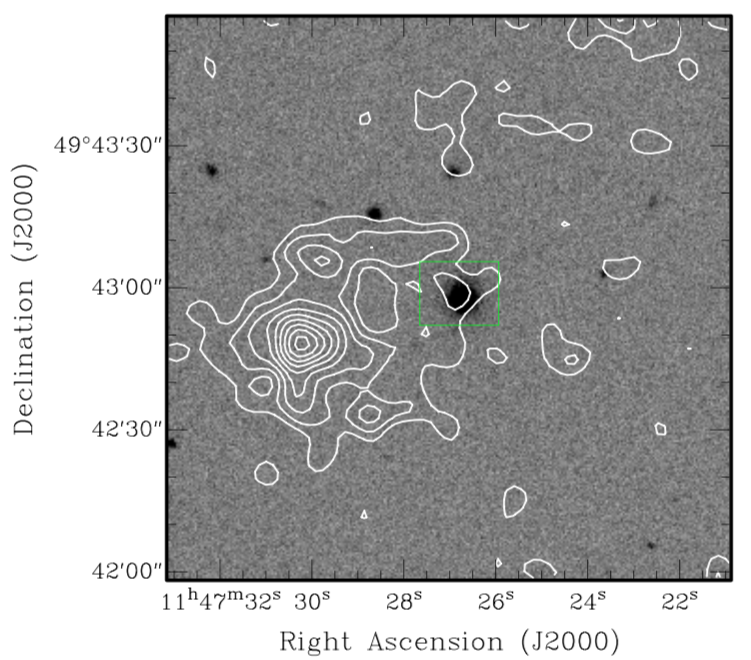}}
  \hfill
\subfloat[{\bf NGC 5905} (thin 0.15 mJy beam$^{-1}\times(1,2,3,4,5)$, thick 1 mJy beam$^{-1}\times(1,2,3,4)$)\label{fig:tde3}]
  {\includegraphics[width=.315\linewidth]{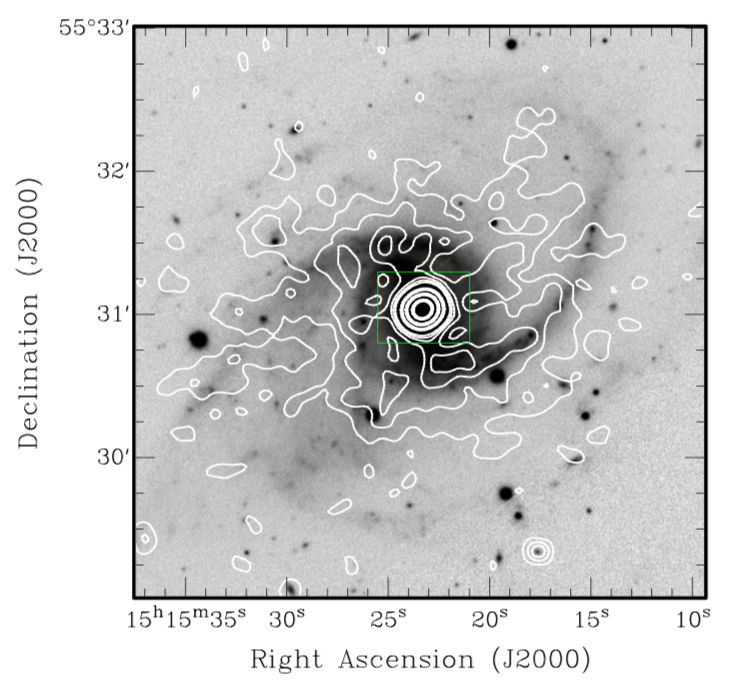}}
\caption{144 MHz radio contours of TDE host galaxies overlaid upon the optical r-band SDSS images. The contour levels are provided in parentheses after the TDE name. The green box indicates location of TDE host galaxy.} 
\label{fig:tde}
\end{figure}

The LoTSS2 survey using the LOFAR provides 144 MHz images covering about 27\% of the northern sky at a median rms of 75 $\mu$Jy beam$^{-1}$ \citep{2022A&A...659A...1S}. The rms in different regions can vary due to variations in de-convolution errors and the sky noise. The LOFAR observations were carried out during 2014 to 2022. In the present release of the LoTSS2, 23 TDEs are covered in the survey images. The timings of the LOFAR detections for these TDEs are between 9029 days and 57 days from the reported day of the TDE transient, wherever known. As some X-ray sources discovered in the X-ray surveys were later identified as TDEs, in those cases the actual epochs of the TDEs remain unknown. We carried out a visual inspection of all 23 fields in the LoTSS2 images. The radio emission is detected from three TDE host galaxies. These images are presented in Fig.~\ref{fig:tde}. The individual objects are discussed below. 
\\
\subsection{AT2018iih} 
AT2018iih was discovered as a transient by the ATLAS facility on November 9, 2018 \citep{2018TNSTR1750....1T} and later classified as a TDE using the the Zwicky Transient Facility \citep{2021ApJ...908....4V}. It is hosted in a dwarf galaxy at $z\sim0.212$ having a blue magnitude of $\sim18$.  The 144 MHz LOFAR radio emission is diffuse and covers several galaxies in the vicinity (see Fig.\ref{fig:tde1}). The epoch of the LOFAR radio observation is 97 days after the discovery of the TDE. The host galaxy appears to be in a multi-galaxy environment and most likely interacting with other galaxies. The total radio flux at 144 MHz is estimated to be $\sim6$ mJy. The radio morphology appears similar to the radio emission in a combined halo formed in a compact group of interacting galaxies.  

\subsection{RBS 1032} 
The RBS 1032 was first detected as a soft and luminous X-ray source in the {\it ROSAT} Bright Survey (RBS) in 1990. Later, it was suggested to be a TDE candidate based on multiple X-ray observations with {\it ROSAT} and {\it XMM-Newton}, and constraining light-curve decay models \citep{2014ApJ...792L..29M}. It is hosted in a dwarf galaxy at $z\sim0.026$. The epoch of the LOFAR radio observation is at least 23 years after the TDE event. The radio morphology appears similar to the radio halo emission in galaxy clusters and groups (see Fig.\ref{fig:tde2}). However, surprisingly no other bright galaxy is seen within the radio extent. There is also no associated diffuse X-ray emission in the {\it ROSAT} and {\it XMM-Newton} images. The radio emission is connected to the TDE host galaxy but not centered on it. This radio morphology at the moment remains unexplained and is highly intriguing in absence of bright optical sources and diffuse X-ray emission. The total radio flux is estimated to be $\sim4$ mJy.

\subsection{ NGC 5905 (RBS 1475)} 
RBS 1475 was also first detected as a soft and luminous X-ray source by {\it ROSAT} and later identified as a TDE candidate based on the X-ray light-curve from multiple X-ray observations at different epochs \citep{1996A&A...309L..35B}. It is hosted in a nearby ($z\sim0.011$) spiral galaxy NGC 5905. The epoch of the LOFAR radio observation is at least 24 years after the TDE event. The 144 MHz morphology is typical of a spiral galaxy showing diffuse emission associated with the the disk and the spiral arms (see Fig. \ref{fig:tde3}). It has a central bright radio source indicating that the nucleus is active. The radio flux of the central source is estimated to be $\sim39$ mJy and that of the diffuse emission is estimated to be $\sim100$ mJy. In a previous attempt with the GMRT, total 1.4 GHz flux associated with the central source was estimated as $\sim9.5$ mJy with no significant disc radio emission \citep{2015Ap&SS.357...32R}. 

\subsection{Sw 1644+57 (non-detection)} 
We also report here an important non-detection at 144 MHz with 3$\sigma$ limit on the flux as $\sim0.3$ mJy for a well-studied TDE Sw 1644+57. The epoch of the LOFAR observation is 2806 days after its discovery. It was first discovered by {\it Swift} and classified as a Gamma Ray Burst GRB110328A \citep{2011GCN.11823....1C}, but later identified as a TDE \citep{2011Sci...333..203B}. Fading radio emission associated with the TDE has been detected previously. Using the radio analysis based on the self-absorption modeling of radio flux at GHz frequencies \citep{2021ApJ...908....4V}, a radio flux much below 0.1 mJy at the LOFAR frequency is predicted. Therefore a non-detection at 144 MHz is consistent here with the previous models.

\bibliography{citations.bib}

\end{document}